\documentclass[fleqn,usenatbib]{mnras}

\usepackage{newtxtext,newtxmath,verbatim}

\usepackage[T1]{fontenc}

\DeclareRobustCommand{\VAN}[3]{#2}
\let\VANthebibliography\thebibliography
\def\thebibliography{\DeclareRobustCommand{\VAN}[3]{##3}\VANthebibliography}

\usepackage{graphicx,float}	
\usepackage{amsmath}	

\usepackage{caption}

\newcommand{\vatot}{v_{\rm{A}}}

\newcommand{\Me}{M_{\rm re}}
\newcommand{\Mr}{M_{\rm r}}
\newcommand{\Be}{v_{\rm e}}

\newcommand{\cost}[1]{\cos^{#1}\theta}
\newcommand{\cosp}[1]{\cos^{#1}\Omega}
\newcommand{\sint}[1]{\sin^{#1}\theta}
\newcommand{\sinp}[1]{\sin^{#1}\Omega}
\newcommand{\vfm}{v_{\rm{f}\perp}}

\title[The Kelvin-Helmholtz instability in relativistic magnetized symmetric flows]{Linear analysis of the Kelvin-Helmholtz instability in relativistic magnetized symmetric flows}

\author[A. Chow et al.]{Anthony Chow$^{1}$\thanks{E-mail: kc3058@columbia.edu},
Michael E. Rowan$^{2}$, 
Lorenzo Sironi$^{1}$,
Jordy Davelaar$^{3,1}$,
 Gianluigi Bodo$^{4}$,
  Ramesh Narayan$^{5}$
\\
$^{1}$Department of Astronomy and Columbia Astrophysics Laboratory, Columbia University, New York, NY 10027, USA\\
$^{2}$Advanced Micro Devices, Inc., Santa Clara, CA, USA\\
$^{3}$Center for Computational Astrophysics, Flatiron Institute, 162 Fifth Avenue, New York, NY 10010, USA\\
$^{4}$INAF Osservatorio Astrofisico di Torino, Via Osservatorio, 20, 10025 Pino Torinese TO, Italy\\
$^{5}$Harvard-Smithsonian Center for Astrophysics, 60 Garden Street, Cambridge, MA 02138, USA
}

\date{Accepted XXX. Received YYY; in original form ZZZ}

\pubyear{2015}

\begin{document}
\label{firstpage}
\pagerange{\pageref{firstpage}--\pageref{lastpage}}
\maketitle

\begin{abstract}
We study the linear stability of a planar interface separating two fluids in relative motion, focusing on the symmetric configuration where the two fluids have the same properties (density, temperature, magnetic field strength, and direction). 
We consider the most general case with arbitrary sound speed $c_{\rm s}$, Alfv\'en speed $v_{\rm A}$, and magnetic field orientation. For the instability associated with the fast mode, we find that the lower bound of unstable shear velocities is set by the requirement that the  projection of the  velocity onto the fluid-frame wavevector is larger than the projection of the Alfv\'en speed onto the same direction, i.e.,  shear should overcome the effect of magnetic tension. 
In the frame where the two fluids move in opposite directions with equal speed $v$, the upper bound of unstable velocities corresponds to an effective relativistic Mach number $\Me \equiv v/\vfm \sqrt{(1-\vfm^2)/(1-v^2)} \cos\theta=\sqrt{2}$, where $\vfm=[\vatot^2+c_{\rm s}^2(1-\vatot^2)]^{1/2}$ is the fast  speed assuming a magnetic field perpendicular to the wavevector (here, all velocities are in units of the speed of light), and $\theta$ is the laboratory-frame angle between the flow velocity and the wavevector projection onto the shear interface. Our results have implications for shear flows in the magnetospheres of neutron stars and black holes --- both for single objects and for merging binaries --- where the Alfv\'en speed may approach the speed of light. 
\end{abstract}




\section{Introduction}
The Kelvin-Helmholtz instability (KHI) \citep{Helmholtz, Kelvin} --- growing in between two fluids in relative motion --- is one of the most common instabilities in space and astrophysical flows.
Since the pioneering works of \citet{chandrasekhar_1961}, the linear theory of the KHI has been investigated under a variety of conditions \citep{turland_scheuer_1976,blandford_pringle_1976,ferrari_trussoni_zaninetti_1980,pu_kivelson_1983, kivelson_zu-yin_1984, bodo_mignone_rosner_2004, osmanov_mignone_massaglia_bodo_ferrari_2008,Blumen_75,Ferrari_78, Sharma_98, 
Prajapati_10,Sobacchi_18, Berlok_19,rowan_phd, Hamlin_Newman_2013, Bodo_Mamtsashvili_Rossi_2013, Bodo_Mamtsashvili_Rossi_2016,Bodo_Mamtsashvili_Rossi_2019, Pimentel_Lora-Clavijo_2019}, depending on whether the two fluids have comparable or different properties (respectively, ``symmetric'' or ``asymmetric'' configuration), whether the flow is incompressible or compressible, and whether or not the fluids are magnetized. Most earlier works  assumed that the relative velocity between the two fluids is non-relativistic.

In \cite{chow_Davelaar_Rowan_Lorenzo_2023}, we studied the linear properties of the KHI for the asymmetric conditions expected at the boundaries of relativistic astrophysical jets. We assumed a magnetically-dominated jet and a gas-pressure-dominated wind, derived the most general form of the dispersion relation, and provided an analytical approximation of its solution for wind sound speeds much smaller than the jet Alfv\'{e}n speed, as appropriate for realistic astrophysical jets.

Although asymmetric conditions are more common in astrophysical systems, it is still interesting to consider the (potentially simpler) symmetric case in which the two fluids have the same properties (sound speed, magnetic field strength, and orientation). Most previous works on the KHI in the relativistic regime focused on the symmetric case. For relativistic hydrodynamic systems (i.e., with no magnetic fields), \cite{bodo_mignone_rosner_2004} obtained an analytical expression for the instability growth rate and derived an absolute upper bound on the range of unstable shear velocities. For symmetric relativistic magnetohydrodynamical flows, \cite{osmanov_mignone_massaglia_bodo_ferrari_2008} considered the case of magnetic field aligned with the direction of relative motion.

In this paper, we study the most general symmetric configuration with arbitrary sound speed, Alfv\'en speed, and magnetic field orientation. We relax the restriction of a flow-aligned magnetic field adopted by \cite{osmanov_mignone_massaglia_bodo_ferrari_2008} and consider the most general scenario where both the unstable wavevector and the magnetic field can have arbitrary directions. We derive the most general form of the dispersion relation and present several analytical results in the astrophysically-relevant regime where magnetic pressure dominates over gas pressure. 

The rest of the paper is organized as follows: \autoref{sec:setup} describes the setup of our study; \autoref{sec:dispersion_relation} derives the dispersion relation for the most general case, having arbitrary Alfv\'{e}n speed and sound speed; \autoref{sec:unstable_boundary} anticipates the range of unstable velocities, and provides a physical interpretation for the lower bound of the unstable region; \autoref{sec:magnetic_dominate} focuses on magnetically-dominated flows (i.e., with magnetic pressure much larger than  gas pressure) and derives analytical results for the growth rate and the range of unstable velocities; \autoref{sec:general_case} discusses  results obtained for the general case in which the gas pressure cannot be neglected. We conclude in \autoref{sec:conclusions}, where we also discuss some astrophysical implications.

\section{Setup} \label{sec:setup}
\begin{figure}
	\centering	\includegraphics[width=0.5\textwidth]{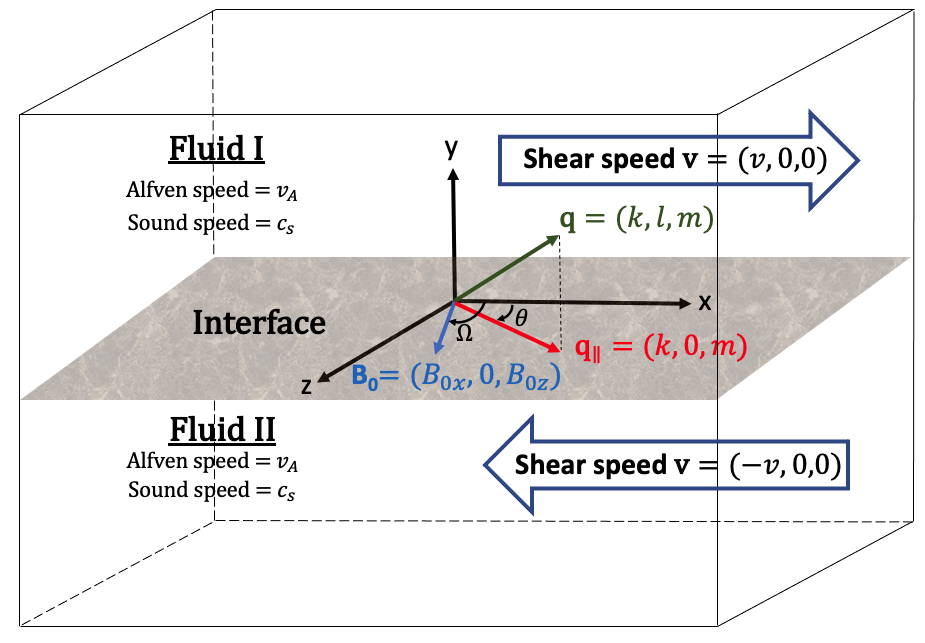}
	\caption{Schematic diagram of the interface separating two fluids in relative motion. The interface (grey color) is located in the $x-z$ plane. Above and below the interface are fluids I and II with the same Alfv\'{e}n speed $\vatot$ and sound speed $c_{\rm s}$. $\mathbf{q}_\parallel$ is the projection of wavevector $\mathbf{q}$ onto the interface. We consider the system in the laboratory frame where fluids I and II have shear velocities $+v\hat{x}$ and $-v\hat{x}$, respectively. $\mathbf{B}_0$ is the magnetic field
vector in the fluid rest frame. $\theta$ is the angle between $\mathbf{q}_\parallel$ and $\hat{x}$ (in the laboratory frame) while $\Omega$ is the angle between
$\mathbf{B}_0$ and $\hat{x}$ (in the fluid frame).} \label{fig:schematic}
\end{figure}

In this work, we consider a magnetized planar vortex sheet in the $x$--$z$ plane at $y=0$, as shown in \autoref{fig:schematic}. The magnetic field in fluid I ($y>0$) and II ($y<0$) is given by $\mathbf{B}_{0}=(B_{{\rm 0x}},0,B_{{\rm 0z}})$, where the $x$ component of the magnetic field, $B_{{\rm 0x}}$, is along the flow direction and the $z$ component, $B_{{\rm 0z}}$, is perpendicular to the flow direction. The magnetic field $\mathbf{B}_0$ is defined in the fluid rest frame. We solve the system in the laboratory frame where fluid I moves with velocity $\mathbf{v}=+v\hat{x}$ and fluid II moves with velocity $\mathbf{v}=-v\hat{x}$. We use subscripts `+' and `-' to denote fluid I or II. We adopt Gaussian units such that $c=4\pi=1$ and define all velocities (i.e., the shear velocity, the sound speed and the Alfv\'en speed) in units of the speed of light $c$.

We describe the flow with the equations of ideal relativistic magnetohydrodynamics (RMHD) \citep[e.g.,][]{anile_1990, mignone_mattia_bodo_2018,rowan_phd}:
\begin{subequations}\label{eq:RMHD_equations}
\begin{align}
    & \frac{\partial (\rho \gamma)}{\partial t}+\nabla \cdot (\rho \gamma \mathbf{v})=0 \\
    &\frac{\partial}{\partial t} (w\gamma^2\mathbf{v}) + \nabla\cdot(w\gamma^2\mathbf{vv})+\nabla p = \rho_{\rm e} \mathbf{E}+\mathbf{J}\times \mathbf{B} \label{eq:momentum_eqn}\\
    & \frac{\partial \mathbf{B}}{\partial t} + \nabla\times\mathbf{E}=0 \label{eq:faraday_eqn}\\
    & \frac{\partial \mathbf{E}}{\partial t}-\nabla \times\mathbf{B}=-\mathbf{J} \label{eq:ampere_eqn}\\
    & \frac{\partial}{\partial t}(w\gamma^2-p)+\nabla\cdot (w \gamma^2 \mathbf{v})=\mathbf{J}\cdot\mathbf{E}
\end{align}
\end{subequations}
supplemented with the divergence constraints
\begin{align}
    \nabla \cdot \mathbf{E}=\rho_{\rm e}, \quad \nabla \cdot \mathbf{B}=0~.
\end{align}
{Here, $\rho$ is the rest mass density, $\rho_{\rm e}$ the charge density, $\mathbf{J}$ the current density vector, $\mathbf{v}$ the fluid velocity vector, $\gamma$ the Lorentz factor, $\mathbf{B}$ the magnetic field vector, $\mathbf{E}$ the electric field vector, $w$ the enthalpy density and $p$ the gas pressure. For an ideal gas with constant adiabatic index $\Gamma$, the enthalpy can be written as $w=\rho+ \Gamma p/(\Gamma-1)$.}

The sound speed \citep[see e.g.][]{mignone_mattia_bodo_2018} and the Alfv\'{e}n speed, both defined in the fluid rest frame, can be written as 
\begin{align}
    c_{s}=\sqrt{\frac{w_{0}-\rho_{0}(\partial w_{0}/\partial \rho_{0})} {(\partial w_{0}/\partial p_{0})-1}\frac{1}{w_{0}}}=\sqrt{\Gamma\frac{p_{0}}{w_{0}}}~
\end{align}
and
\begin{align}
{\bf{v}}_{\rm A}=\frac{{\bf B}_0}{\sqrt{w_0+B_{{\rm 0x}}^2+B_{{\rm 0z}}^2}}~,
\end{align}
where ``0'' subscripts indicate  variables in the equilibrium state. We define $v_{\rm A}$ as the magnitude of the Alfv\'{e}n speed.

\section{General Dispersion relation} \label{sec:dispersion_relation}
The dispersion relation of the surface wave at the interface can be obtained from the dispersion relation of the body waves of each fluid (Alfv\'en, slow, fast modes) together with the pressure balance and displacement matching across the interface \citep[e.g.,][]{bodo_mignone_rosner_2004}. The dispersion relation of the body wave can be found by linearization of \autoref{eq:RMHD_equations} in the fluid rest frame, by substituting the following perturbed variables  
\begin{subequations}\label{eq:pertubed_varaibles}
\begin{align}
    {\rho} &\approx{\rho_0}+{\rho_1} ~,\\
    {p}&\approx {p}_0+{p}_1 ~,\\
    {\mathbf{v}}&\approx\mathbf{0}+{\mathbf{v}}_1 ~,\\
    {\mathbf{E}}&\approx\mathbf{0}+{\mathbf{E}}_1 ~,\\
    \mathbf{B}&\approx\mathbf{B}_0+\mathbf{B}_1~, \\
    {\rho}_{\rm e} &\approx 0 + {\rho}_{ \rm e1}~,
\end{align}
\end{subequations}
where ``$1$'' subscripts indicate first-order perturbed variables, all defined in the fluid rest frame. The perturbed electric field in the fluid is given by ${\mathbf{E}}_1=-{\mathbf{v}}_1 \times \mathbf{B}_0$ in the ideal MHD limit, and the current density is eliminated from \autoref{eq:RMHD_equations} by using \autoref{eq:ampere_eqn}. We consider perturbed variables ${X}_1$ to be plane waves, of the form ${X}_1 \propto e^{i(\tilde{\mathbf{q}}\cdot \mathbf{x}-\tilde{\omega} t)}$ where $\tilde{\mathbf{q}}=(\tilde{k},\tilde{l},\tilde{m})$ is the complex wavevector and $\tilde{\omega}$ is the complex angular frequency, both defined in the rest frame of each fluid (we use overtilde to denote rest-frame frequencies and wavevectors). By solving the linearized system of RMHD equations, the dispersion relation of the body waves can then be written as
\begin{align} \label{eq:dm_body_wave}
\tilde{\omega}_\pm (\tilde{\omega}_\pm^2 &- 
     \tilde{q}_\pm^2 \vatot^2\cos^2\tilde{\varphi}_\pm) [\tilde{\omega}_\pm^4-(\vatot^2+c_{\rm s}^2(1-\vatot^2))\tilde{q}_\pm^2\tilde{\omega}_\pm^2 \nonumber \\
    &+(\tilde{q}_\pm^2-\tilde{\omega}_\pm^2)c_{\rm s}^2\vatot^2\tilde{q}_\pm^2\cos^2\tilde{\varphi}_\pm]= 0~,
\end{align}
where 
\begin{align}
 \cos\tilde{\varphi}_\pm = \frac{\tilde{k}_\pm B_{{\rm 0x}}+\tilde{m}_\pm B_{{\rm 0z}}}{\sqrt{\tilde{k}_\pm^2+\tilde{l}_\pm^2+\tilde{m}_\pm^2} \sqrt{B_{{\rm 0x}}^2+B_{{\rm 0z}}^2}}~,
 \end{align}
is the cosine of the angle between the magnetic field and the fluid-frame wavevector $\tilde{\mathbf{q}}$. The second term of \autoref{eq:dm_body_wave} (in round brackets) corresponds to Alfv\'{e}n waves, which lead to stable solutions. The third term in \autoref{eq:dm_body_wave}  (in square brackets) has solutions
\begin{align} \label{eq:magnetosonic_speed}
    \frac{\tilde{\omega}_\pm^2}{\tilde{q}_\pm^2} &= \frac{1}{2} \left(\vatot^2+c_{\rm s}^2(1-\vatot^2) + \vatot^2 c_{\rm s}^2\cos^2\tilde{\varphi}_\pm \nonumber \right.\\
&\quad \left. \pm \sqrt{[\vatot^2+c_{\rm s}^2(1-\vatot^2\sin^2\tilde{\varphi}_\pm)]^2-4\vatot^2 c_{\rm s}^2\cos^2\tilde{\varphi}_\pm} \right )~,
\end{align}
which correspond to fast and slow magnetosonic modes (plus and minus sign, respectively), which are potentially unstable \citep{osmanov_mignone_massaglia_bodo_ferrari_2008}. By setting $\tilde{\varphi}_\pm=\pi/2$ in the fast mode, we can define a frame-independent quantity
\begin{align}
\label{eq:fast}
    \vfm = \sqrt{\vatot^2 + c_{\rm s}^2(1-\vatot^2)}~,
\end{align}
which is the fast magnetosonic speed in the case that the fluid-frame wavevector and the magnetic field are orthogonal. Hereafter, we will use $\vfm$ as defined in \autoref{eq:fast} to parameterize our results, even in the general case when the  wavevector and the magnetic field are not orthogonal.

Notice that the transverse components of the wavenumber, $l$ and $m$, are Lorentz invariant. By applying the following transformations 
\begin{align}
\tilde{\omega}_\pm = \gamma (\omega\mp k v), \quad \tilde{k}_\pm = \gamma (k\mp\omega v), \quad \tilde{l}_\pm =  l_\pm, \quad \tilde{m}_\pm = m,
\label{eq:ltran}
\end{align}
to \autoref{eq:dm_body_wave}, we can construct the Lorentz invariant ratio $l_+^2/l_-^2$:
\begin{align} \label{eq:lPluslMinus2_general}
    \frac{l_+^2}{l_-^2}=f(B_{{\rm 0x}}/B_{{\rm 0z}},v, c_s,\vatot, k, m, \omega)~,
\end{align}
where the function $f$ can be written explicitly as a function of the parameters indicated in parenthesis. 
An independent way of obtaining $l_+/l_-$ is to solve the linearized RMHD equations, Eqs.~(\ref{eq:RMHD_equations}), together with two additional constraints. First, we enforce the first-order pressure balance between the two sides of the interface:
\begin{align}
    p_{1+} \!+\!  B_{{\rm 0x}}B_{\rm 1x,+}\!+\!B_{{\rm 0z}}B_{\rm 1z,+} 
    \!=\!p_{1-} \!+\!B_{{\rm 0x}}B_{\rm 1x,-}\!+\!B_{{\rm 0z}}B_{\rm 1z,-}~. 
\end{align}
Second, the fluid displacements in the $y$ direction must match at the interface. Since the Lagrangian derivative of the displacement is equal to the transverse velocity $v_{1\rm y}$ of the fluid, matching the displacements is equivalent to
\begin{align}
    \frac{v_{\rm 1y,-}}{\gamma(\omega+kv)}=\frac{v_{\rm 1y,+}}{\gamma(\omega-kv)}~.
\end{align}
This results in an independent solution for the ratio $l_+/l_-$:
\begin{align} \label{eq:lPluslMinus_general}
    \frac{l_+}{l_-}=g(B_{{\rm 0x}}/B_{{\rm 0z}}, v,c_{\rm s},\vatot, k, m, \omega)~.
\end{align}
where the function $g$ can be written as a function of the parameters indicated in parenthesis.

We can finally equate \autoref{eq:lPluslMinus2_general} and the square of \autoref{eq:lPluslMinus_general} to obtain the dispersion relation in an implicit form:
\begin{align} \label{eq:general_dm}
    f=g^2~.
\end{align}
We can then rearrange \autoref{eq:general_dm} by introducing the dimensionless ratios
\begin{align} \label{eq:symmetric_parameters}
    &\phi = \frac{\omega}{\vfm \sqrt{k^2+m^2}}, \quad M=\frac{v}{\vfm},
\end{align}
where $M$ is the flow fast Mach number, i.e., the flow speed in units of the fast magnetosonic speed defined in \autoref{eq:fast}. With this parameterization, the dispersion relation in \autoref{eq:general_dm} takes the form
\begin{align} \label{eq:general_quartic_poly}
    \left( \frac{\phi}{M}\right )\left[ c_0 + c_1\left(\frac{\phi}{M} \right)^2+c_2\left(\frac{\phi}{M} \right)^4+c_3\left(\frac{\phi}{M} \right)^6+c_4\left(\frac{\phi}{M} \right)^8\right ]=0,
\end{align}
where $c_0,c_1,c_2,c_3$ and $c_4$ are constants with respect to $\phi$. Note that Im$(\phi)>0$ implies that the amplitude of the perturbation grows exponentially, i.e., the system becomes unstable.
Since the factor $\phi/M$ in front of \autoref{eq:general_quartic_poly} corresponds to a stable mode, \autoref{eq:general_quartic_poly} is effectively a quartic equation in $(\phi/M)^2$ and can be solved analytically \citep{weisstein}, giving a total of eight (generally, complex) roots. However, not all of them may be acceptable. First, since we have introduced spurious roots when squaring it, i.e., not all solutions will satisfy \autoref{eq:lPluslMinus_general}. Also, only outgoing waves should be retained by the Sommerfeld radiation condition \citep{sommerfeld_1912}. This requires accepting only those roots that satisfy the conditions:
\begin{align}
    &\text{Im}(l_+)>0 \\
    &\text{Im}(l_-)<0.
\end{align} 

In the rest of the paper, we present solutions to the dispersion relation. We first focus on the special case of magnetically-dominated flows, i.e.,  $c_{\rm s}\ll v_{\rm A}$, and then we address the most general case. We employ the following parameters:
\begin{align}
M_{\rm r}=M\frac{\sqrt{1-\vfm^2}}{\sqrt{1-v^2}}, \quad \cos\Omega = \frac{B_{{\rm 0x}}}{\sqrt{B_{{\rm 0x}}^2+B_{{\rm 0z}}^2}}, \quad \cos\theta = \frac{k}{\sqrt{k^2+m^2}}, 
\end{align}
where $M_{\rm r}$ is a ``relativistic'' Mach number defined with the spatial part of the 4-velocity, as in \citet{osmanov_mignone_massaglia_bodo_ferrari_2008}, $\Omega$ is the angle between the $+\hat{x}$ direction and the magnetic field {(in the fluid rest frame)}, and $\theta$ is the angle {(measured in the laboratory frame)} between $+\hat{x}$ and the wavevector projection onto the $y=0$ interface. It is also convenient to absorb the dependence on $\cos\theta$ by defining an effective Mach number $\Me$ and an effective flow velocity $\Be$ similarly to the purely hydrodynamic case discussed by \citet{bodo_mignone_rosner_2004}:
\begin{align} \label{eq:symeter_me_be}
\Me=M_{\rm r}\cos\theta=\frac{v}{\vfm}{}\frac{\sqrt{1-\vfm^2}}{\sqrt{1-v^2}}\cos\theta,\quad \Be=v \cos\theta.
\end{align}
In the following, we usually write our solution for $\phi$ in terms of $\Me$ and $\Be$ (clearly, when $\cos\theta=1$, we have $M_{\rm re}=M_{\rm{r}}$ and $v_{\rm e}=v$). For $c_{\rm s} =0$, in some cases, it will be convenient to use $\vatot$ instead of $\Be$. Assuming $c_{\rm s} =0$, the two are related by the following identity
\begin{equation}
\Be = \frac{\vatot \Me \cost{}}{\sqrt{\cost{2}-\vatot^2\cost{2}+\Me^2 \vatot^2}}~.
\end{equation}
\section{Boundaries of the unstable region} \label{sec:unstable_boundary}
We anticipate here some general remarks on the range of unstable velocities $v$ (or equivalently, of $\Me$), which will be confirmed below in several specific cases. For the upper bound, we expect the system to be unconditionally stable if $\Me\geq \sqrt{2}$. This bound has already been discussed in the purely hydrodynamic case $\vatot=0$ by \citet{bodo_mignone_rosner_2004}, where $\Me$ was defined with the sound speed. Here, we find that the same bound holds, as long as $\Me$ is defined with the fast magnetosonic speed as in \autoref{eq:symeter_me_be}.

A physically-insightful lower bound can be found in the case of instability associated with the fast mode. As shown below, this case (as opposed to the case of instability associated with the slow mode) is the most important one since it leads to faster growth rates, and it is the only one that persists for magnetically-dominated flows with $c_{\rm s}=0$. As we show in \autoref{sec:phase}, unstable modes associated with the fast magnetosonic wave have vanishing phase speeds (so, zero real part of $\omega$). The boundaries of the unstable region will also have, by construction, a vanishing growth rate (so, zero imaginary part of $\omega$). This implies that $\tilde{k}_\pm=\gamma k$ (see \autoref{eq:ltran}) so that, in the fluid comoving frame, the angle $\tilde{\theta}$ between the ``projected wavevector'' $\tilde{\bf q}_\parallel$ (i.e., the projection of $\tilde{\bf q}$ onto the $x-z$ interface) and the $\hat{x}$ direction of the shear velocity satisfies
\begin{align}
\label{eq:kvect}
\gamma\tan\tilde{\theta}=\tan {\theta}~.
\end{align}
We expect the system to be unstable only if the projection of the shear velocity onto the direction of $\tilde{\bf{q}}_\parallel$ is larger than the projection of the Alfv\'en speed onto the same direction. Equivalently,
\begin{equation}
v\cos\tilde{\theta}>v_{\rm A}\cos(\Omega-\tilde{\theta})~.
\end{equation}
In other words, instability happens when the shear can overcome the effect of magnetic tension (for particular cases of this general criterion, see \cite{pu_kivelson_1983, Roychoudhury_Lovelace_1986}). Using \autoref{eq:kvect}, this can be written as a function of lab-frame quantities as 
\begin{equation}
\label{eq:general}
\frac{v}
{\vatot}>\cos\Omega+\frac{\sin\Omega\,\tan\theta}{\gamma}~.
\end{equation}
We demonstrate below that this successfully describes the lower bound of unstable $v$ in the case of fast-mode instability for all $c_{\rm s}$, $\vatot$, and field orientations.

\section{Magnetically-dominated flows} \label{sec:magnetic_dominate}
Although the dispersion relation in \autoref{eq:general_quartic_poly} for the most general symmetric shear flows is analytically solvable, the solution is too complex for further analysis of the instability region. However, in certain regimes of our parameter space, the solution simplifies and is physically interpretable. This section considers the special case where the flow is magnetically dominated ($c_{\rm s}\ll\vatot$). We then  assume the cold plasma limit ($c_{\rm s}=0$) and the solution is analytically tractable and allows for physical insight.

For magnetically-dominated flows, we assume $c_{\rm s}=0$ so that $v_{\rm{f}\perp}=\vatot$ and the dispersion relation (\autoref{eq:general_quartic_poly}) is reduced to a quadratic equation in $(\phi/M)^2$ with solutions: 
\begin{align} \label{eq:general_soln_cs_0}
    \frac{\phi^2}{M^2} &= \frac{\mu_1 \pm \sqrt{\mu_2}}{\mu_3}
\end{align}
where $\mu_0,\mu_1$ and $\mu_2$ are given in \autoref{sec:explicit_mu} for completeness.

We now consider a few special cases. First, we consider the case where the projected wavevector is parallel to the shear flow ($\cos\theta=1$) for arbitrary $\vatot$ and $\cos\Omega$. Then, we retain the dependence on $\cos\theta$ and $\vatot$, but consider the specific cases of in-plane or out-of-plane fields ($B_{\rm 0z}=0$ or $B_{\rm 0x}=0$, respectively; equivalently, $\cos\Omega=1$ or $\cos\Omega=0$, respectively). Finally, we retain the angular dependence on $\cos\theta$ and $\cos\Omega$, but we consider the extreme limit $\vatot\rightarrow1$. These cases are summarized in \autoref{tab:tab1}.

\begin{table} \caption{Magnetically-dominated cases.} \label{tab:tab1} \begin{tabular}{ccccc} \hline
$\vatot$ & $\cos\theta$ & $\cos\Omega$ & Growth Rate & Unstable Range\\
\hline
\hline
Arbitrary & 1 & Arbitrary & \autoref{eq:symmetric_cs0_costheta1_soln} & \autoref{eq:KHI_boundary_cosphi}\\
Arbitrary & Arbitrary & 1 & \autoref{eq:inplane_cs0_soln_2}& \autoref{eq:in_plane_cold_boundary}\\
Arbitrary & Arbitrary & 0 & \autoref{eq:symmetric_outplane_soln} & \autoref{eq:out_plane_cold_boundary} \\
1 & Arbitrary & Arbitrary & \autoref{eq:general_cs0_vA_1} & \autoref{eq:general_cs0_va1_boundary}\\
\hline
\end{tabular} \end{table}

\subsection{Magnetically-dominated flows with projected wavevector along the shear velocity}
We now consider the case where the projected wavevector is parallel to the shear flow ($\cos\theta=1$, in which case $M_{\rm re}=M_{\rm{r}}$ and $v_{\rm e}=v$), where the solution of \autoref{eq:general_soln_cs_0} can be written as 
\begin{align} 
    \frac{\phi^2}{M^2} &=\frac{\kappa_1\pm\sqrt{\kappa_2}}{\kappa_3}, \label{eq:symmetric_cs0_costheta1_soln}
\end{align}
where
\begin{align}
\kappa_1&=\Mr^4 \vatot^2 (1 - \vatot^2\cos^2\Omega)^2 + 
 \Mr^2 (1 - \vatot^2) (1 -  \vatot^2\cos^2\Omega)^2\nonumber \\
 &+ (1 - \vatot^2) (1 -  \vatot^2\cos^4\Omega) \nonumber \\
\kappa_2&= (1 - 
   \vatot^2\cos^2\Omega )^2 [(1 - \cos^2\Omega)^2 (1 - 
      \vatot^2)^2 \nonumber \\
   &+4 \Mr^2 (1 - \cos^2\Omega \vatot^2)^2 (1 - \vatot^2 + \Mr^2\vatot^2)]\nonumber\\
  \kappa_3&=\Mr^2 [1 - \vatot^2 + 
   \Mr^2 \vatot^2 (1 - \vatot^2\cos^2\Omega )] [1 + (1 - 
      2 \cos^2\Omega) \vatot^2 \nonumber \\
      &+ \Mr^2 \vatot^2 (1 - \vatot^2\cos^2\Omega)]/(1 - \vatot^2 + \Mr^2 \vatot^2). \nonumber
\end{align}
Since $\kappa_2$ and $\kappa_3$ are always positive, the condition for instability can be found from the inequality $\kappa_1^2<\kappa_2$. In this case, the solution is purely imaginary, and the instability range in $\Mr$ reads

\begin{align} \label{eq:KHI_boundary_cosphi}
    \frac{\cos\Omega \sqrt{1-\vatot^2}}{\sqrt{1-\vatot^2\cos^2\Omega}} < \Mr <\frac{\sqrt{2-\cos^2\Omega-\vatot^2\cos^2\Omega}}{\sqrt{1-\vatot^2\cos^2\Omega}}~.
\end{align}
\autoref{eq:KHI_boundary_cosphi} completely characterizes the boundary of the instability region for cold flows when $\cos\theta=1$. Note that the left inequality in \autoref{eq:KHI_boundary_cosphi} can be cast as $\vatot \cos\Omega < v$,
which implies that the interface is unstable only when the shear speed is greater than the projection of the Alfv\'{e}n speed onto the flow direction (in this case, the latter coincides with the direction of the projected wavevector), in agreement with \autoref{eq:general} for ${\theta}=0$. 

By setting $\cos\Omega=0$ in \autoref{eq:KHI_boundary_cosphi}  (i.e., magnetic field perpendicular to the shear flow), the instability region attains its maximum range. Similarly to the purely hydrodynamic case (but with the fast speed now replacing the sound speed in $\Me$), we find
\begin{align} \label{eq:max_bound_phi}
    0<\Mr<\sqrt{2}~. 
\end{align} 

\subsection{Magnetically-dominated flows with in-plane magnetic field} \label{subsec:inplane_cold_plasma}
When the magnetic field is parallel to the  flow velocity, i.e., $\cos\Omega=1$, \autoref{eq:general_soln_cs_0} can be rewritten as:
\begin{align}
    \frac{\phi^2}{M^2}&=\frac{\nu_1\pm \sqrt{\nu_2}}{\nu_3} \cost{2} \label{eq:inplane_cs0_soln_2}
\end{align}
where
\begin{align} 
    &\nu_1=\Me^4+\Be^2+\Me^2(1-\Be^2), \nonumber \\
    &\nu_2=\sin^4\theta(\Me^2+\Be^2)^2+4\Me^4[\Be^2+\Me^2(1-\Be^2)], \nonumber \\
    &\nu_3=(\Me^2+\Be^2)^2. \nonumber
\end{align}
Since $\nu_2$ and $\nu_3$ are always positive, the condition for instability can be found from the inequality $\nu_1^2<\nu_2$. In this case, the solution is purely imaginary, and the instability range in $\Me$ is
\begin{align} \label{eq:in_plane_cold_boundary}
    \cos\theta <\Me  < \sqrt{2-\cos^2\theta}~, 
\end{align}
where the case $\cos\theta=0$ should be excluded since it yields a vanishing growth rate.
\autoref{eq:in_plane_cold_boundary} completely characterizes the boundary of the instability region for cold flows when $\cos\Omega=1$. {Note that the left inequality in \autoref{eq:in_plane_cold_boundary} can be cast as $ \vatot < v$, implying that the interface is unstable only when the flow is super-Alfv\'{e}nic. This is consistent with the general expression in \autoref{eq:general} for the case $\Omega=0$ considered here.
} 

By setting $\cos\theta\rightarrow 0$ in \autoref{eq:in_plane_cold_boundary}, the instability region attains its maximum range
\begin{align} \label{eq:max_bound_inplane}
    0<\Me<\sqrt{2}~.
\end{align}

\subsection{Magnetically-dominated flows with out-of-plane magnetic field} \label{subsec:outplane_cold_plasma}
We now consider the case of the magnetic field perpendicular to the shearing flow, i.e., $\cos\Omega=0$. Since the instability boundaries depend explicitly on $\vatot$, we choose to express the solution in terms of $\vatot$ instead of $\Be$. By setting $\Omega=\pi/2$ in \autoref{eq:general_soln_cs_0}, 
we obtain the solution for $\phi$ in the form
\begin{align} 
    \frac{\phi^2}{M^2} &=\frac{\lambda_1\pm\sqrt{\lambda_2}}{\lambda_3} \cos^2\theta~, \label{eq:symmetric_outplane_soln}
\end{align}
where
\begin{align*}
    \lambda_1&=   \Me^2 (1 + \Me^2) \vatot^2 - \cost{4} \vatot^2 (1 - \vatot^2) \\
    &+ \cost{2} [1 - \vatot^4 + \Me^2 (1 - 2 \vatot^2)]\\
    \lambda_2&=\cost{8} (1\! -\! \vatot^2)^2 + 
 \Me^4 \vatot^2 (4 \cost{2}\! + \!4 \Me^2 \sint{2} \vatot^2 + 
    \sint{4} \vatot^6) \\
    &+ 2 \cost{2} \Me^2 (1 \!- \!\vatot^2) [2 \Me^2 \sint{2} \vatot^2 \!+\!
    \cost{4} \vatot^4\! +\! \cost{2} (2\! - \!\vatot^4)]\\
    \lambda_3&=\Me^4 \vatot^2\! +\! \cost{2} \Me^2 (1 \!+\! \vatot^2)~.
\end{align*}
Since $\lambda_2$ and $\lambda_3$ are always positive, the criterion for unstable modes can be derived from the inequality $\lambda_1^2 < \lambda_2$. The solution is purely imaginary, and the instability range in $\Me$ is
\begin{align} \label{eq:out_plane_cold_boundary}
   \sqrt{(1-\cos^2\theta)(1-\vatot^2)} < \Me < \sqrt{1+\cos^2\theta +\vatot^2(1-\cos^2\theta)}~,
\end{align}
{where the case $\cos\theta=0$ should be excluded since it yields a vanishing growth rate.} \autoref{eq:out_plane_cold_boundary} completely characterizes the boundary of the instability region for cold flows when $\cos\Omega=0$. {Note that the left inequality of \autoref{eq:out_plane_cold_boundary} can be cast as $\vatot \tan\theta<\gamma v$. Once again, this is consistent with \autoref{eq:general} for the case $\Omega=\pi/2$ considered here.} 

By setting $\cos\theta=1$, the instability region attains its maximum range
\begin{align}
    \label{eq:max_bound_outplane}
    0<\Me<\sqrt{2}~,
\end{align}
as we had already derived in \autoref{eq:max_bound_phi} under similar assumptions ($\cos\theta=1$ and $\cos\Omega=0$). \\

\subsection{Flows with Alfv\'en speed approaching the speed of light}
The system remains unstable in the extreme limit $\vatot\rightarrow 1$. In this case, \autoref{eq:general_soln_cs_0} reduces to
\begin{align} \label{eq:general_cs0_vA_1}
    \frac{\phi^2}{M^2}=\frac{\eta_1^2-\eta_2^2}{\eta_3}~,
\end{align}
where
\begin{align*}
    \eta_1&=\gamma\cost{}\cosp{}+\sint{}\sinp{}~,\\
    \eta_2&=\gamma v\cost{}~, \\
     \eta_3&=\gamma^2 v^2 (1-v^2\cosp{2})~.
\end{align*}
Since $\eta_3$ is always positive, the criterion for unstable modes can be derived from the inequality $\eta_1^2<\eta_2^2$. In this case, the solution is purely imaginary, and the instability range is
\begin{align}\label{eq:general_cs0_va1_boundary}
    \cosp{} + \frac{\sinp{}\tan\theta}{\gamma}< v <1~.
\end{align}
The lower limit coincides with \autoref{eq:general} for $\vatot=1$.

In the special case of out-of-plane fields, i.e., $\cos\Omega=0$, the instability growth rate is
\begin{align}
    \frac{\phi^2}{M^2}=\frac{\sin^2\theta-v^2}{v^2}~,
\end{align}
whereas the instability range can be cast as 
\begin{align}
{\sin\theta< v < 1} ~.
\end{align}

\section{General Case} \label{sec:general_case}
 In this section, we discuss the dependence of the growth rate on $c_{\rm s}$ and $\vatot$. The general solution for arbitrary $c_{\rm s}$ and $\vatot$ is analytically available but too lengthy to be presented, so we focus on numerical results.
\subsection{Comparison to the hydrodynamic case}\label{sec:hydro}
When the unstable modes propagate perpendicular to the magnetic field, we expect magnetic tension to have no effect, and the solution should resemble the hydrodynamic symmetric case discussed by \citet{bodo_mignone_rosner_2004}, {but now with the fast magnetosonic speed instead of the sound speed in the definition of $\Me$}. We demonstrate this by setting $\cos\Omega=0$ and $\cos\theta=1$ in the general dispersion relation (\autoref{eq:general_quartic_poly}) and the solution can be written as 
\begin{align}
\label{eq:bodo}
    \frac{\phi^2}{M^2}= \frac{1+\Mr^2-v^2\pm\sqrt{4\Mr^2(1-v^2)+(1+v^2)^2}}{\Mr^2+2v^2}
\end{align}
which is essentially the same as  
 \equationautorefname{ 15} in \cite{bodo_mignone_rosner_2004} for relativistic hydrodynamic flows when the projected wavevector is parallel to the shear flow. We conclude that even though the system is magnetized, in the case when $\cos\Omega=0$ and $\cos\theta=1$, the instability behaves similarly to the hydrodynamic case. Here, the magnetic field provides pressure but not tension. We remind that the Mach numbers in \autoref{eq:bodo} are defined using $v_{\rm{f}\perp}$, and so we allow for an arbitrary degree of gas versus magnetic contributions to the pressure (in contrast to \cite{bodo_mignone_rosner_2004}, which assumed $\vatot=0$). In analogy to what found by \citet{bodo_mignone_rosner_2004}, the instability range is 
\begin{equation}
    0<M_{\rm r}<\sqrt{2}~.
\end{equation}
 
\begin{figure}
	\centering
	\includegraphics[width=0.5\textwidth]{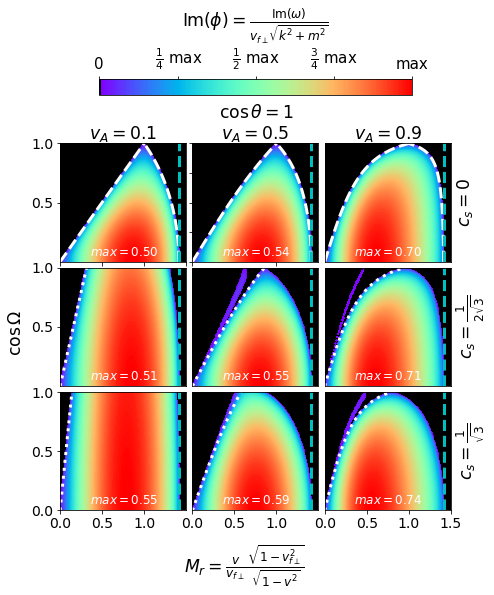}
	\caption{Dependence of the instability growth rate Im($\phi$) on $M_r$ and $\Omega$ when $\theta=0$ (i.e., wavevector along the shear flow), for three choices of $\vatot$ and $c_{\rm s}$. Panels in the first, second, and third columns represent $\vatot=0.1,0.5$ and $0.9$. Panels in the first, second, and third rows represent $c_{\rm s}=0,1/2\sqrt{3},1/\sqrt{3}$. In all the panels, Im($\phi$) is then normalized to its maximum value, which is quoted in the panels themselves. The dashed cyan vertical lines represent the common upper bound $\Me=\sqrt{2}$ across all panels. The dashed white lines in the top three panels represent the exact boundaries of the instability region when $c_{\rm s}=0$ and $\cos \theta=1$, see \autoref{eq:KHI_boundary_cosphi}. The dotted white lines in the middle and bottom rows represent the lower bound of the fast-mode unstable region imposed by magnetic tension, as in \autoref{eq:general}.} \label{fig:general_inplane_phi_3x3} \label{fig:cold_plasma_limit_phi}
\end{figure}
\begin{figure}
    \centering	
    \includegraphics[width=0.5\textwidth]{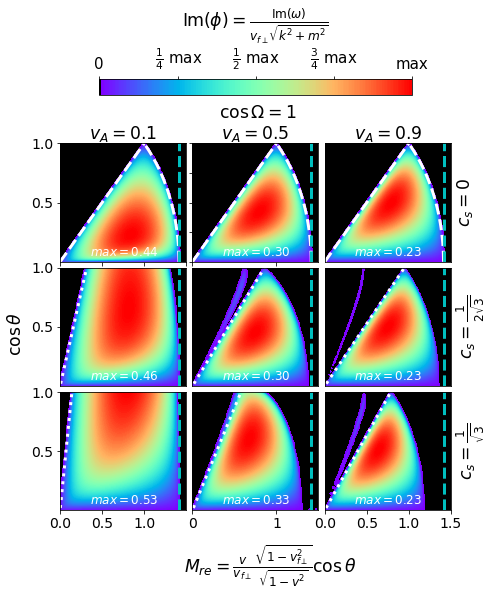}
    \caption{Dependence of the instability growth rate Im($\phi$) on $\Me$ and $\theta$ when $\Omega=0$ (i.e., magnetic field along the flow direction), for three choices of $\vatot$ and $c_{\rm s}$.  The dashed cyan vertical lines represent the common upper bound $\Me=\sqrt{2}$ across all panels. The dashed white lines in the top three panels represent the exact boundaries of the instability region when $c_{\rm s}=0$ and $\cos \Omega=1$, see \autoref{eq:in_plane_cold_boundary}.  See the caption of \autoref{fig:general_inplane_3x3} for further details.} \label{fig:general_inplane_3x3}
\end{figure}
\begin{figure}
	\centering
	\includegraphics[width=0.5\textwidth]{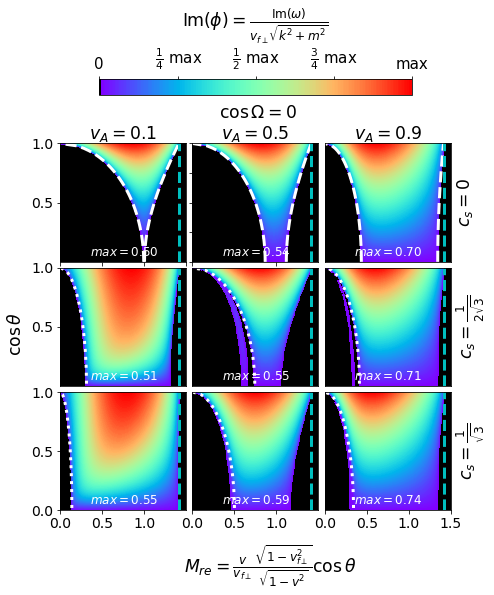}
	\caption{Dependence of the instability growth rate Im($\phi$) on $\Me$ and $\theta$ when $\Omega=\pi/2$ (i.e., magnetic field perpendicular to the flow direction), for three choices of $\vatot$ and $c_{\rm s}$. The dashed cyan vertical lines represent the common upper bound $\Me=\sqrt{2}$ across all panels. The dashed white lines in the top three panels represent the exact boundaries of the instability region when $c_{\rm s}=0$ and $\cos \Omega=0$, see \autoref{eq:out_plane_cold_boundary}. See the caption of \autoref{fig:general_inplane_3x3} for further details.} \label{fig:general_outplane_3x3}
\end{figure}
\subsection{General case} 
We now consider the problem in its full generality, allowing for arbitrary $c_{\rm s}$, $\vatot$, $\theta$, and $\Omega$. 
\autoref{fig:general_inplane_phi_3x3} shows the dependence of the instability growth rate Im($\phi$) on $\Me$ and $\cos\Omega$ when the wavevector is parallel  to the shear flow ($\cos\theta=1$, in which case $\Me=M_r$). \autoref{fig:general_inplane_3x3} and \autoref{fig:general_outplane_3x3} show the dependence of the instability growth rate on $\Me$ and $\cos\theta$ when the orientation of the magnetic field is parallel ($\cos\Omega=1$) and perpendicular ($\cos\Omega=0$)  to the shear flow, respectively. The case in \autoref{fig:general_inplane_3x3} is the same as in \cite{osmanov_mignone_massaglia_bodo_ferrari_2008}. In all three figures, we consider different combinations of $\vatot$ and $c_{\rm s}$. The panels in the left, middle, and right columns correspond to $\vatot=0.1, 0.5$ and $0.9$, respectively; the panels in the top, middle and bottom rows correspond to $c_{\rm s}=0, 1/\sqrt{12}$ and $1/\sqrt{3}$ respectively. The figures plot the imaginary part of $\phi$ (i.e., the growth rate). We present the real part of the solution  (i.e., the phase speed) in \autoref{sec:phase}. 

In all figures, we identify two separate instability regions in the middle-middle, middle-right, and bottom-right panels. The narrow purple instability stripes are associated with the slow-mode magnetosonic wave, while the large instability regions correspond to the fast-mode wave,\footnote{More precisely, the narrow purple stripes correspond to slow waves in one fluid and fast waves in the other fluid; the large instability region corresponds to fast waves in both fluids.} see \cite{osmanov_mignone_massaglia_bodo_ferrari_2008}. For $c_{\rm s}=0$ (top three panels in all figures), only the fast-mode solution survives. The middle-left, bottom-left, and bottom-middle panels do not have visible slow-mode instability bands, but further analysis reveals that they exist near the left boundaries of the main instability zones whenever $c_{\rm s}\neq0$. 

In \autoref{sec:phase}, we show that unstable modes associated with the slow wave have nonzero phase speed. In contrast, unstable modes associated with the fast magnetosonic wave have vanishing phase speed. As discussed in \autoref{sec:unstable_boundary}, the lower bound of unstable velocities --- regarding the fast-mode instability --- should obey \autoref{eq:general}. This is depicted as a dotted white line in the figures (for the top row, \autoref{eq:general} corresponds to the left branch of the dashed lines), confirming that \autoref{eq:general} successfully describes the lower bound of unstable $\Me$ of the fast-mode instability.

For the gas pressure dominated case with $c_{\rm s}=1/\sqrt{3}$ and $\vatot=0.1$ (bottom-left panels in Figures \ref{fig:general_inplane_phi_3x3}-\ref{fig:general_outplane_3x3}), we expect the orientation of the magnetic field (i.e., $\cos\Omega$) to have little effect on the instability growth rate, as indeed confirmed by the bottom-left panel of \autoref{fig:general_inplane_phi_3x3}. In the hydrodynamic case, \citet{bodo_mignone_rosner_2004} found that the growth rate peaks at $\cos\theta=1$, it vanishes for $\cos\theta=0$, and that the range   of unstable Mach numbers is $0<\Me<\sqrt{2}$ for all $\cos\theta\neq0$. This is in agreement with the near-hydro case displayed in the 
 bottom-left panels of \autoref{fig:general_inplane_3x3} and \autoref{fig:general_outplane_3x3}.

For magnetically-dominated flows, we first consider the top row of \autoref{fig:general_inplane_3x3} where the plasma is cold ($c_{\rm s}=0$), and the magnetic field is along the shear velocity ($\cos\Omega=1$). When the wavevector is perpendicular to the flow ($\cos\theta=0$), the flow is stable, regardless of $\vatot$. When the wavevector is parallel to the magnetic field ($\cos\theta=1$ in this case), the effect of magnetic tension in suppressing the instability is the greatest, and the flow is also stable. Since the system is stable when $\cos\theta$ is either 0 or 1, the maximum growth rate is attained at intermediate values of $\cos\theta$.
The boundary of the instability region is fully characterized by \autoref{eq:in_plane_cold_boundary}, which is depicted as white dashed lines in \autoref{fig:general_inplane_3x3}. The left boundary of the unstable region corresponds to $\vatot<v$, indicating that the system is unstable only when the shear can overcome magnetic tension. The upper bound of the instability region is determined by the right inequality in \autoref{eq:in_plane_cold_boundary}. For $\cos\theta\rightarrow0$, magnetic pressure dominates over tension, and the range of unstable Mach numbers approaches the one of the hydrodynamic
case (compare with the bottom-left panel in \autoref{fig:general_inplane_3x3}).

The magnetically-dominated flows with $\cos\Omega=0$, i.e., field orthogonal to the shear velocity (top row in \autoref{fig:general_outplane_3x3}) show  instability profiles which are vastly different than the case in \autoref{fig:general_inplane_3x3}. At $\cos\theta=1$, the magnetic field is perpendicular to the wavevector, parallel to the shear flow. Since magnetic tension plays no role, both the instability growth rate and the unstable Mach number range at $\cos\theta=1$ resemble those of the gas-pressure dominated case (bottom left panel of \autoref{fig:general_outplane_3x3}; see also our discussion in \autoref{sec:hydro}). When $\cos\theta \rightarrow 0$, the wavevector aligns more and more closely to the magnetic field, and magnetic tension gets stronger. Therefore, the instability growth rate gets suppressed when $\cos\theta \rightarrow 0$, and the instability range shrinks. It is interesting to note that at intermediate $\cos\theta$ between 0 and 1, the range of unstable Mach numbers widens for higher $\vatot$. This has a simple explanation: for fixed $\cos\theta$ and fixed $\Me$, increasing $\vatot$ implies an increase in $v$. In the fluid frame, the projected wavevector of the fast-mode instability forms an angle $\tilde{\theta}$ with respect to the shear velocity such that $\gamma\tan\tilde{\theta}=\tan {\theta}$, see \autoref{eq:kvect}. Therefore, with increasing $v$, the fluid-frame wavevector becomes more aligned with the flow and nearly orthogonal to the magnetic field. This implies that at fixed $\cos\theta$ and increasing $\vatot$, the system can more easily overcome the effect of magnetic tension.

For magnetically-dominated flows with arbitrary orientation of the magnetic field and fixed $\cos\theta=1$ (top row in \autoref{fig:general_inplane_phi_3x3}), the instability profiles at $\cos\Omega=1$ and $\cos\Omega=0$ obviously coincide with those of \autoref{fig:general_inplane_3x3} and \autoref{fig:general_outplane_3x3} for $\cos\theta=1$. The white dashed boundaries of the instability region are described by \autoref{eq:KHI_boundary_cosphi}, where the left inequality can be cast as $\vatot\cos\Omega<v$, indicating that the interface is unstable only when the shear speed is greater than the projection of the Alfvén speed onto the flow direction (which, in this case, coincides with the direction of the projected wavevector).

For the general case where both $\vatot$ and $c_{\rm s}$ are nonzero (middle row of Figures \ref{fig:general_inplane_phi_3x3}-\ref{fig:general_outplane_3x3}), the instability behavior can be understood as an intermediate case between the gas-pressure-dominated case and the magnetically-dominated case.

Finally, we consider the global boundary of the instability region in the general case. Although the general dispersion relation in \autoref{eq:general_quartic_poly} is analytically solvable, the explicit expression of the unstable growth rate is too complicated to derive an instability boundary in terms of $\Me$ analytically. However, the instability criteria that we derived for magnetically-dominated flows in \autoref{sec:magnetic_dominate} allow us to argue for a global instability boundary in the general case. In the top three panels of Figures \ref{fig:general_inplane_phi_3x3}-\ref{fig:general_outplane_3x3}, the white dashed lines correspond to the instability boundaries described by Equation \ref{eq:KHI_boundary_cosphi}, \ref{eq:in_plane_cold_boundary} and \ref{eq:out_plane_cold_boundary} respectively. In each case, as well as in the purely hydrodynamic case, the maximum instability range is bounded below by 0 and above by $\sqrt{2}$, as shown in Equation \ref{eq:max_bound_phi}, \ref{eq:max_bound_inplane} and \ref{eq:max_bound_outplane}. The lines $\Me=\sqrt{2}$ (the cyan dashed lines  in all panels) appear to provide an upper limit to the unstable region in all cases. Therefore, we argue that the maximum instability range in $\Me$ for general symmetric flows is
\begin{align}
    0<\Me<\sqrt{2}~,
\end{align}
i.e., that flows with $\Me\geq\sqrt{2}$ are unconditionally stable.

\section{Discussion and Conclusions}\label{sec:conclusions}
We have studied the linear stability properties of the KHI for relativistic, symmetric, magnetized flows. We considered arbitrary sound speeds, Alfv\'en speeds, and magnetic field orientations and derived the most general form of the dispersion relation.

Our results show that, for  $c_{\rm s} \neq 0$, there are two distinct unstable regions, corresponding to the slow and fast modes of the magnetosonic wave, consistent with the findings in \cite{osmanov_mignone_massaglia_bodo_ferrari_2008}. For the instability associated with the fast mode, which leads to greater growth rates, we obtained a lower bound on the range of unstable shear velocities, see \autoref{eq:general}. This lower bound reveals a necessary condition for the KHI growth, namely that the projection of the shear velocity $v$ onto the direction of $\tilde{\bf q}_\parallel$ (which, itself, is the fluid-frame wavevector projected onto the interface) should be larger than the projection of the Alfv\'{e}n speed onto the same direction. In other words, for the fast mode to become unstable, the free energy of the shear must overcome magnetic tension. Concerning the upper bound of unstable shear velocities, we found that the system is unconditionally stable if $\Me \geq \sqrt{2}$, regardless of $c_{\rm s}$, $\vatot$ or magnetic field direction. Here,  $\Me$ is an effective Mach number defined with the fast magnetosonic speed, see \autoref{eq:symeter_me_be}.  

By considering the cold plasma limit $c_{\rm s}=0$ (equivalently, the limit of magnetically-dominated plasmas), we derived an analytically tractable dispersion relation and computed the instability growth rate and the range of unstable shear velocities for several special cases (see \autoref{tab:tab1}). In the limiting case of Alfv\'en speed approaching the speed of light, we found that the system is still unstable as long as the shear speed is greater than the lower bound in \autoref{eq:general_cs0_va1_boundary}. 

Although most shear flows in the Universe are expected to have an asymmetric configuration --- e.g., the boundaries of AGN jets \citep{chow_Davelaar_Rowan_Lorenzo_2023} ---  the results obtained here for a symmetric setup in the magnetically-dominated regime $\vatot\gg c_{\rm s}$ may have important implications for the magnetospheres of neutron stars and black holes --- both for single objects and for merging binaries. 

Binary neutron stars in the latest stages of inspiral are unlikely to be tidally locked \citep{bildsten_92}. Due to the orbital motion, the interface between the two magnetospheres can be modeled as a magnetically-dominated shear layer. This configuration is analogous to the symmetric setup we adopted in this work. In the final stages of the inspiral, when the surfaces of the two neutron stars come into contact, the KHI and ensuing turbulence have also been invoked to govern rapid magnetic field amplification via dynamo processes (e.g., \citet{Price_Rossowog_2006,zrake_13,Kiuchi+15}). 

For single objects, shear motions in the azimuthal direction (equivalently, discontinuities in the profile of the angular velocity) have been revealed by fully kinetic and fluid simulations, both for black hole magnetospheres \citep[e.g., Figure 3 in][]{elmellah_22} and for pulsar magnetospheres \citep[e.g., Figure 4 in][]{Timokhin_2006}. In addition, in active pulsars, a discontinuity in the poloidal flow velocity is expected to occur near the star, between the open field lines (loaded with radially streaming pairs) and the closed magnetosphere.

We conclude with a few caveats. The plane-parallel approach we employed is applicable only if the width of the shear layer is small compared to other length scales of the system (e.g., for binary neutron stars, their separation). Also, we have assumed that the fluids on the two sides of the interface have the same sound speed, magnetic field strength, and orientation. In the magnetically-dominated regime, pressure balance requires the magnetic field strength on both sides to be the same. However, the field direction will generally be different. In fact,  in nearly all of the astrophysical applications discussed above, the surface of velocity shear also presents a significant magnetic shear. Our simplifying assumptions will be relaxed in a future work.

\section*{Acknowledgements}
We are grateful to H. Hakobyan,  J. N\"attil\"a, and especially J. Mahlmann for useful discussions. L.S. acknowledges support from the Cottrell Scholars Award, and DoE Early Career Award DE-SC0023015. L.S. and J.D. acknowledge support from NSF AST-2108201. This research was facilitated by Multimessenger Plasma Physics Center (MPPC), NSF grant PHY-2206609. J.D. is supported by a Joint Columbia University/Flatiron Research Fellowship, research at the Flatiron Institute is supported by the Simons Foundation.

\section*{Data Availability}
No new data were generated or analysed in support of this research.

\providecommand{\noopsort}[1]{}





\onecolumn
\appendix
\section{General solution for magnetically-dominated flows} \label{sec:explicit_mu}
For magnetically-dominated flows, we assume $c_{\rm s}=0$ so that $v_{\rm{f}\perp}=\vatot$ and the dispersion relation (\autoref{eq:general_quartic_poly}) is reduced to a quadratic equation in $(\phi/M)^2$ with solutions: 
\begin{align} 
    \frac{\phi^2}{M^2} &= \frac{\mu_1 \pm \sqrt{\mu_2}}{\mu_3}
\end{align}
where
\begin{align}
    \mu_1 &= \cost{2}\Me^4 (1 +\Me^2) \gamma^5 - 
 \cost{3} \cosp{2} (\Me^2) [
   \cosp{2} - 2 (1 +\Me^2)] \sint{} (1-\gamma^2) \gamma^4 \nonumber \\ 
   & \quad \quad + 
 \cost{8} \sinp{2} (1-\gamma^2)^2 (\gamma^3) [
   \cosp{2} (1 + \gamma^2) - \gamma^2] - 
 \cost{7} \cosp{2} \sint{} (1-\gamma^2)^2 (\gamma^2) [
   \cosp{2} (1 + 3 \gamma^2) - 3 \gamma^2] \nonumber \\
   & \quad \quad + 
 \cost{4} \Me^2 (1-\gamma^2) (\gamma^3) [
   \cosp{4} (2 \sint{2} - 1) - 1 - 2 \gamma^2 + 
    \Me^2 (1 - 2 \gamma^2) + 
    \cosp{2} (1 + 2 (1 + \Me^2) \gamma^2)] \nonumber \\
    &\quad\quad - 
 \cost{6} (1-\gamma^2) (\gamma^3) [\gamma^4 - 1 + 
    \Me^2 (1 - 3 \gamma^2 + \gamma^4) + 
    \cosp{2} (1 + \gamma^2 - 2 \gamma^4 + 
       \Me^2 (1 + 4 \gamma^2 - 2 \gamma^4)) \nonumber\\
       &\quad \quad+ 
    \cosp{4} ((1-\gamma^2) (2 \sint{2} - \gamma^2) + 
       \Me^2 (\gamma^4 - 2 - \gamma^2))] - 
 \cost{5} \cosp{2} \sint{} (1-\gamma^2) (\gamma^2) [
   (1-\gamma^2) (1 + \sinp{2} \gamma^2) \nonumber \\
   & \quad\quad+ 
    \Me^2 (6 \gamma^2 - 1 - 2 \gamma^4 + 
       \cosp{2} (-1 - 5 \gamma^2 + 2 \gamma^4))] \\
 \mu_2 &= \cost{}^4 \gamma^4 [\cost{} \cosp{2} \sint{} \
(1-\gamma^2) + \Me^2 \gamma - 
    \cost{2} \sinp{4} (1-\gamma^2) \gamma]^2 \Me^4 (\cosp{4} + 4 \Me^2) \gamma^4 \nonumber \\
    &\quad\quad- 
 4 \cost{} \cosp{2} \Me^4 \sint{} \gamma^3 (\cosp{2} - 2 + 2 \gamma^2) + 
 \cost{8} (\cosp{2} - \gamma^2 + \gamma^4 - 
    \cosp{2} \gamma^4)^2 \nonumber \\
    &\quad\quad+ 
 4 \cost{7} \cosp{2} \sint{} (1-\gamma^2)^2 \gamma [
   \cosp{2} (1 + \gamma^2) - \gamma^2] + 
 4 \cost{5} \cosp{2} \sint{} (1-\gamma^2) \gamma [
   (1-\gamma^2)^2 + \Me^2 (5 \gamma^2 - 2 \gamma^4 - 1) \nonumber \\
   &\quad\quad- 
    \cosp{2} (2 - (3 - 6 \Me^2) \gamma^2 + (1 - 
          2 \Me^2) \gamma^4)] + 
 4 \cost{3} \cosp{2} \Me^2 \sint{} \gamma [
   (1-\gamma^2) (2 \sinp{4} \gamma^4 - 
       1 + (-1 + 4 \cosp{2}) \gamma^2) \nonumber\\
       &\quad\quad+ 
    \Me^2 (1 - 4 \gamma^2 + 2 \gamma^4 + 
       \cosp{2} (1 + \gamma^2))] - 
 2 \cost{2} \Me^2 (\gamma^2) [(\cosp{4}) [
     2 - 3 \gamma^2 + \gamma^4 + 
      \Me^2 (1 - 2 \sint{2} + \gamma^2)] \nonumber \\
      & \quad\quad- 
    2 \Me^2 (1-\gamma^2) (\Me^2 - 2 \gamma^2 - 1) + 
    \cosp{2} (\Me^2 (4 \gamma^4 - 1 - 4 \gamma^2) - 
       (1-\gamma^2)^2)] + 2 \cost{6} (1-\gamma^2) [(\gamma^2) [
     \Me^2 (1 - 4 \gamma^2 + 2 \gamma^4) \nonumber\\
     &\quad\quad- (1-\gamma^2)^2] + (
     \cosp{2}) [
     1 + \gamma^2 - 4 \gamma^4 + 2 \gamma^6 + 
      \Me^2 (2 \gamma^2 - 1 + 7 \gamma^4 - 4 \gamma^6)] + 
    \cosp{4} ((1-\gamma^2) [\gamma^4 - 
         2 - (1 - 2 \sint{2}) \gamma^2] \nonumber \\
         &\quad\quad+ 
       \Me^2 (1 - 4 \gamma^2 - 3 \gamma^4 + 2 \gamma^6))] + (
  \cost{4}) [
  (1-\gamma^2)^4 + \Me^4 (1 - 8 \gamma^2 + 16 \gamma^4 - 8 \gamma^6) + 
   2 \Me^2 (6 \gamma^2 - 1 - 7 \gamma^4 + 2 \gamma^8)  \nonumber \\
   &\quad\quad+(
    \cosp{4}) [
    \Me^4 (1 + \gamma^2)^2 + (2 - 3 \gamma^2 + \gamma^4)^2 - 
     4 \Me^2 (1-\gamma^2) (1 - 2 (2 - \sint{2}) \gamma^2 + \gamma^6)] + 
   2 (\cosp{2}) [
     (1-\gamma^2)^3 (\gamma^2 - 2)\nonumber \\
     &\quad\quad + 
      \Me^4 (3 \gamma^2 - 8 \gamma^4 + 4 \gamma^6 - 1) + 
      \Me^2 (3 - 14 \gamma^2 + 13 \gamma^4 + 2 \gamma^6 - 
         4 \gamma^8)]] \\
    \mu_3 &= [\Me^2 \gamma^2 - 
   \cost{2} (1-\gamma^2) (\cosp{2} (1-\gamma^2) + \gamma^2)] [
 2 \cost{4} \sinp{2} (1-\gamma^2)^2 \gamma - 
  2 \cost{3} \cosp{2} \sint{} (1-\gamma^2)^2 \nonumber \\
  &\quad\quad + 
  2 \cost{} \cosp{2} \Me^2 \sint{} (1-\gamma^2) \gamma^2 + 
  \Me^4 \gamma^3+ 
  \cost{2} \Me^2 \gamma (\gamma^2 - 2 + \gamma^4 + 
     \cosp{2} (1 - \gamma^4))]
\end{align}

\section{Phase speed of the unstable modes}\label{sec:phase}
\begin{figure}
    \captionsetup{width=0.4\textwidth}
	\centering
	\includegraphics[width=0.5\textwidth]{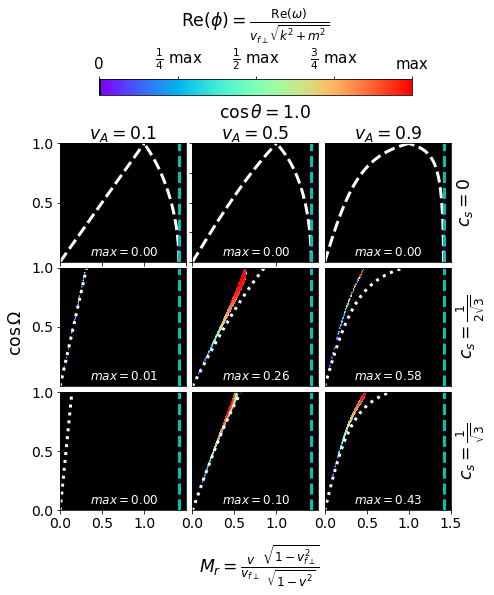}
	\caption{Dependence of the real part of  $\phi$ on $\Me$ and $\Omega$ when $\cos\theta=1$. The corresponding imaginary part of the solution is in  \autoref{fig:general_inplane_phi_3x3}.}
	  \label{fig:general_inplane_phi_3x3_real}
\end{figure}

\begin{figure}
	\centering
	\includegraphics[width=0.5\textwidth]{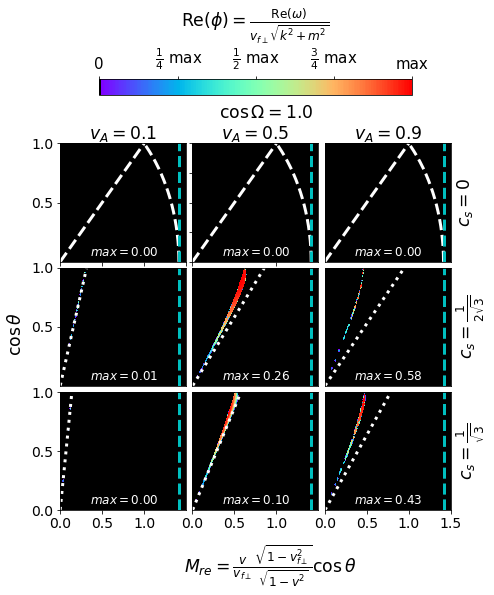}
		\caption{Dependence of the real part of  $\phi$ on $\Me$ and $\theta$ when $\cos\Omega=1$. The corresponding imaginary part of the solution is in \autoref{fig:general_inplane_3x3}.}
	\label{fig:general_inplane_3x3_real}
\end{figure}

\begin{figure}
	\centering
	\includegraphics[width=0.5\textwidth]{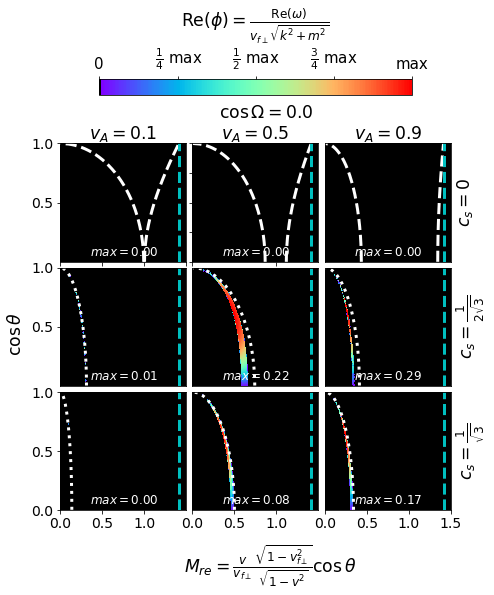}
	\caption{Dependence of the real part of  $\phi$ on $\Me$ and $\theta$ when $\cos\Omega=0$. The corresponding imaginary part is in \autoref{fig:general_outplane_3x3}.}
\label{fig:general_outplane_3x3_real}
\end{figure}

In the main text, we presented three figures showing the dependence of the imaginary part of $\phi$ (i.e., the growth rate) on $\Me$, $\cos\theta$, $\cos\Omega$, $\vatot$ and $c_{\rm s}$. In this Appendix, we provide corresponding plots for the real part (i.e., the phase speed). {In \autoref{sec:unstable_boundary}, we claimed that unstable modes associated with the fast magnetosonic wave have vanishing phase speed. \autoref{fig:general_inplane_phi_3x3_real}, \autoref{fig:general_inplane_3x3_real} and \autoref{fig:general_outplane_3x3_real} plot the real part of $\phi$, whose imaginary part was presented  in \autoref{fig:general_inplane_phi_3x3}, \autoref{fig:general_inplane_3x3} and \autoref{fig:general_outplane_3x3} respectively. To guide the eye, we overplot the boundaries of the unstable region associated with the fast mode (white dashed and dotted lines). In all three figures,  non-zero phase speeds are all located outside of the fast-mode unstable boundary, indicating that the unstable modes associated with the fast wave indeed have vanishing phase speed. In contrast, unstable modes associated with the slow wave (more precisely, these  modes correspond to slow waves in one fluid, and fast waves in the other fluid) have nonzero phase speed.}

\bsp	
\label{lastpage}
\end{document}